# Dependence of Dopant Geometry on Na Concentration in Na$_x$CoO$_2$

M. H. N. Assadi and H. Katayama-Yoshida

*Graduate School of Engineering Science, Osaka University, Osaka 560-8531, Japan*

*assadi@aquarius.mp.es.osaka-u.ac.jp*

Na$_x$CoO$_2$ has recently been the focus of renewed attention as a promising material for high-efficiency thermoelectric systems. Na$_x$CoO$_2$ lattice consists of alternating Na layers and edge-sharing CoO$_6$ octahedra as schematically represented in **Fig. 1**. The electronically frustrated triangular CoO$_2$ layer in Na$_x$CoO$_2$ possesses massive spin entropy which results in large Seebeck coefficient. Additionally, the irregularities in the Na layer at room temperature scatter phonons thus lower the thermal conductivity.[1] For this, Na$_x$CoO$_2$ has an unprecedented advantage over other thermoelectric materials when it comes to favourably adjusting the interdependent parameters of the *figure of merit* (*ZT*) as a phonon-glass electron-crystal system. From a materials engineering viewpoint, controlling Na concentration has been one of the tools to push the *ZT* of Na$_x$CoO$_2$ to higher limits. There is plenty of research relating Na concentration to Na$_x$CoO$_2$'s electrical and crystallographic properties. The other tool to improve the thermoelectric performance is doping. For instance, in the case of Yb, 3% of Yb dopants significantly increased the power factor to $1.5 \times 10^{-3}$ W m$^{-1}$K$^{-2}$, albeit with a trade-off between increased resistivity.[2] Other dopants with similar effects are Mg, Ni, Zn, Eu and etc.[3,4]

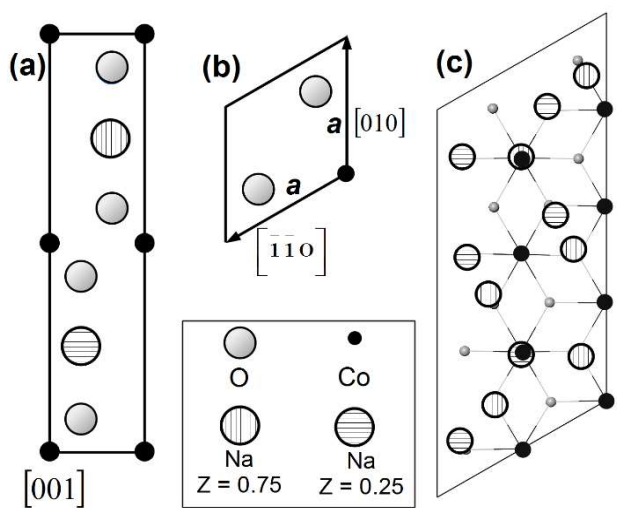

**Fig. 1**. (a) The schematic presentation of the primitive cell and lattice vectors of the NaCoO$_2$ crystal. There are two Na$^+$ ions in each primitive cell, one at Z = 0.25 and the other at Z = 0.75. (b) The position of O atom in the CoO$_2$ layer. (c) The positions of the Na$^+$ ions in the supercell Na$_{0.75}$CoO$_2$ systems preventative of Na pattern for a given Na concentration.

Further progress in the field requires a fundamental atomistic level understanding of electric behaviour as a function of Na concentration in Na$_x$CoO$_2$. One particular concern is the difference in the bonding nature and lattice structures of the CoO$_2$ and Na layers in Na$_x$CoO$_2$.

This may cause different dopants to be stable at different lattice sites as Na concentration varies. Interactions governing dopants' behaviours are strongly dominated by the chemical coordination which requires theoretical probe using quantum many-body techniques. In this work, we investigated the behaviour of Sb dopants in $Na_xCoO_2$ for Na concentrations of $x$ = 0.75, 0.875 and 1.00 by density functional theory. We chose $Na_xCoO_2$ with higher Na concentration of $x > 0.75$ because it has excessively higher thermopower[5] thus it is appealing for practical applications. The rationale for choosing Sb was its exceedingly higher atomic mass than all elements of the host crystal enabling Sb to rattle phonons considerably.

The main findings of our work can be summarised as follow:

(1)     Sb stabilises at its highest oxidation state, 5+, regardless of lattice site that it incorporates into or even Na concentration. This results in substantial electron-hole recombination and reduces the hole concentration of the Sb-doped $Na_xCoO_2$ systems.

(2)     In sodium-rich systems ($x$ = 1 and 0.875), Sb dopant stabilises at the $CoO_2$ layer in the form of $Sb_{Co}^{5+}$. However, the margin of stability of $Sb_{Co}^{5+}$ decreases as Na concentration decreases.

(3)     By varying Na concentration, one can dictate the incorporation lattice site of Sb dopants. Sb incorporation site also correlates with the $Na_xCoO_2$' calculated fundamental band gap which offers a solution for bandgap engineering. This conclusion may be generalised to other positively charged dopants in Na deficient $Na_xCoO_2$.